\documentclass[letter]{ptptex}
\notypesetlogo

\title{
A Phantom does not Result from a Backreaction%
}

\author{
Hisako \textsc{Tanaka}\footnote{E-mail:
  sushia@astr.tohoku.ac.jp}
and 
Toshifumi \textsc{Futamase}\footnote{E-mail:
  tof@astr.tohoku.ac.jp} 
}

\inst{
Astronomical Institute, Tohoku University, 
Sendai 980-8578, Japan
}
\recdate{October 31, 2007}

\abst{
The backreaction of nonlinear inhomogeneities to the cosmic expansion 
is analyzed in the framework of general relativity with a cosmological constant.  
By defining the spatially averaged matter energy density, 
we find that the cosmological constant induces a new type of 
backreaction whose equation of state is 
$p=-4/3 \rho$, where $\rho$ and $p$ are the effective energy density 
and effective pressure of the backreaction in the averaged Friedmann universe.
However, the effective density is negative, and thus it decreases the acceleration caused by the 
cosmological constant.
}

\begin{document}

\maketitle

The observation of the isotropy of the cosmic microwave background radiation (CMBR) 
\cite{WMAP} and large galaxy surveys, such as SDSS\cite{SDSS}, indicate 
that the universe is isotropic and homogeneous over scales of about 100 Mpc. 
However, local matter distributions are highly inhomogeneous. 
This simple observation makes naive use of an isotropic and homogeneous 
model of the universe, namely the Friedmann-Lemaitre-Robertson-Walker (FLRW) model 
questionable, since the solution of the Einstein equations with an averaged 
homogeneous matter distribution does not solve the Einstein equations 
with a realistic matter distribution, because of its nonlinearity.  
Therefore, it is naturally expected that if we average a locally inhomogeneous universe, 
the expansion of the averaged spacetime will be affected somehow by the local inhomogeneities. 
This effect is called the backreaction due to local inhomogeneities, and there have been 
many investigations of this backreaction \cite{
Futamase88,Futamase89, Futamase96,Kasai92,Kasai93,Kasai95,Russ96,Russ97,SA,AK}. 

The problem of backreaction has attracted much attention recently. 
Observations of Type Ia supernovae 
\cite{Riess,Perlmutter} and the CMBR \cite{FGC,BC} 
strongly suggest that the cosmic expansion is accelerating. 
Understanding the source of this accelerated expansion is one of the 
greatest unsolved problems in modern cosmology \cite{Bahcall,PR}. 
Usually the existence of an unknown type of energy (dark energy, 
perhaps a cosmological constant) is assumed for the explanation of this acceleration. 
A possible alternative idea is that 
the energy caused by inhomogeneities leads to additional terms 
in the Friedmann equation, as if dark energy did exist \cite{Rasanen,Kolb05}. 
Although this possibility is unlikely\cite{IW, HS}, it is still important to 
derive a definite conclusion in regard to it and to carefully study its properties. 

In a previous study\cite{KAF} we showed in the case without a 
cosmological constant that the nonlinear backreaction neither accelerates 
nor decelerates the cosmic expansion in a matter-dominated universe and  
that the backreaction behaves as a small positive curvature term. 
This conclusion was obtained by defining the averaged density of matter 
as a conserved quantity in the large comoving volume. 
Note that there are many FLRW spacetimes  
that fit a particular inhomogeneous universe. Our choice 
is motivated by considering how one constructs a FLRW 
universe through actual observations, namely, by defining the evolution of the number density  
of galaxies by comparing with the comoving constant number density.  
It was also shown that the above results do not depend on the definition of 
the averaging procedure, and the results in the Newtonian gauge are consistent with 
those in the comoving synchronous gauge.  

However, the above results were obtained in the case with no cosmological constant, 
and it is not at all clear if the same conclusions apply in the case with  a 
cosmological constant. On the basis of observations, it appears that the existence of the cosmological 
constant is very likely and thus it is worthwhile studying the backreaction 
due to local inhomogeneities in a universe with a more general background. 
This is the purpose of the present study.

We formulate the problem of the backreaction as generally as possible. For this purpose, 
we follow Kasai, Asada and Futamase\cite{KAF} and employ the $3+1$ formulation 
in general relativity. 
We restrict ourselves to the case of 
a vanishing shift vector, i.e., $N_{i}=0$. The line element is written 
\begin{equation} 
ds^{2}= -(Ndt)^{2}+\gamma_{ij}dx^{i}dx^{j}. 
\end{equation}
The unit vector  normal to a hypersurface foliated by $t=const$ is denoted by
 \begin{equation} 
n^{\mu}=\left(\frac{1}{N},0,0,0\right). 
\end{equation}
The extrinsic curvature is defined as 
\begin{equation}
 K^{i}{}_{j}=\frac{1}{2N}\gamma^{ik}\dot{\gamma}_{kj}, 
\end{equation}
where the dot denotes the time derivative. The Einstein equations with a cosmological constant 
are reduced to the Hamiltonian constraint, the momentum constraint and 
(the trace of) the evolution equation as 
\begin{equation} 
{}^{(3)}R+(K^{i}{}_{i})^{2}-K^{i}{}_{j}K^{j}{}_{i}=16\pi GE +2\Lambda,
\end{equation}
\begin{equation}
K^{i}{}_{j|i}-K^{j}{}_{i|j}=8\pi GJ_{i},
\end{equation}
\begin{equation} 
\dot{K}^{i}{}_{i}+NK^{i}{}_{j}K^{j}_{i}-N^{|i}{}_{|i}=-4\pi GN(E+S) +N\Lambda,
\end{equation}
where$ {}^{(3)}R$ is a 3-dimensional Ricci scalar curvature, ${}_{|i}$ 
denotes the 3-dimensional covariant derivative, and the other symbols are defined as follows: 
\begin{equation}
 E=T_{\mu\nu}n^{\mu}n^{\nu}=\frac{1}{N^{2}}T_{00},
\end{equation}
\begin{equation}
J_{i}=-T_{\nu i}n^{\nu}=\frac{1}{N}T_{0i},
\end{equation}
\begin{equation}
 S=T_{ij}\gamma^{ij}.
\end{equation} 
We define the spatial 3-dimensional volume $V$ of a compact domain $D$ in a $t=const$ 
hypersurface as
\begin{equation}
V=\int_{D}\sqrt{\gamma}d^{3}x ,
\end{equation}
where 
\begin{equation}
 \gamma=\det(\gamma_{ij}).
 \end{equation} 
Here, $V$ is considered to be a
 volume sufficiently large that we can assume periodic boundary conditions.

The scale factor $a(t)$ is defined in terms of the volume expansion rate of the universe: 
\begin{equation} 
3\frac{\dot{a}}{a}\equiv\frac{\dot{V}}{V}
=\frac{1}{V}\int_{D}\frac{1}{2}\gamma^{ij}\dot{\gamma}_{ij}\sqrt{\gamma}d^{3}x .
\end{equation}
Now we adopt the average procedure 
\begin{equation} 
 \langle{A}\rangle  \equiv\frac{1}{V}\int_{D}A\sqrt{\gamma}d^{3}x .
\end{equation}
Then, we have 
\begin{equation}
 3\frac{\dot{a}}{a}= \langle{NK^{i}{}_{i}}\rangle.
\end{equation}
Furthermore, we define the deviation from uniform Hubble flow as
\begin{equation}
 V^{i}{}_{j}\equiv NK^{i}{}_{j}-\frac{\dot{a}}{a}\delta^{i}{}_{j} .
\end{equation}
Then we can show $\langle V^{i}{}_{i}\rangle=0$.

By averaging the Einstein equations, we obtain 
\begin{equation}
  \left(\frac{\dot{a}}{a}\right)^2 = \frac{8\pi G}{3}\langle{N^2E}\rangle -
  \frac{1}{6} \langle{N^2\, {}^{\scriptscriptstyle(3)}\!R}\rangle -
  \frac{1}{6} \langle{(V^i_{\ i})^2 - V^i_{\ j} V^j_{\ i}}\rangle + 
  \frac{\Lambda}{3} \langle{N^2}\rangle, 
\end{equation}
\begin{equation}
  \frac{\ddot{a}}{a} =
  -\frac{4\pi G}{3}\langle{N^2(E+S)}\rangle
  + \frac{1}{3} \langle{(V^i_{\ i})^2 - V^i_{\ j} V^j_{\ i}}\rangle
  + \frac{1}{3} \langle{N N^{|i}_{\ |i} + \dot{N} K^i_{\ i}}\rangle
  + \frac{\Lambda}{3} \langle{N^2}\rangle . 
\end{equation}
The geometrical treatment of these equations has studied by Buchert\cite{Buchert00, Buchert06}.

In this paper we approximate the matter of the universe as irrotational dust whose energy-momentum tensor is given by
 \begin{equation}
 T^{\mu\nu}=\rho u^{\mu}u^{\nu},
 \end{equation}
where $u^{\mu}$ is the 4-dimensional velocity of the fluid flow.
 
Up to this point, our equations has been completely general, and a more detailed treatment 
requires explicit expressions for the inhomogeneous universe. In this paper 
we employ the post-Newtonian approximation for the metric to obtain a more concrete 
expression for the backreaction terms, which allows us to obtain a  physical interpretation.

As the background solution, we employ the totally flat universe with a non-vanishing 
cosmological constant. It is then sufficient to consider the first-order 
cosmological post-Newtonian metric, 
\begin{equation}
ds^{2}=- (1+2\phi({\bf x}, t))dt^{2}+a^{2}(1-2\phi({\bf x}, t))\delta_{ij}dx^{i}dx^{j},
\end{equation}
where $\delta_{ij}$ denotes the Kronecker delta. Note that the potential $\phi$ does depend 
on time in general. However, here we restrict ourselves to perturbations  
with scales much smaller than the horizon scale, $L_H$; otherwise the concept 
of averaging becomes unclear. 
This allows us to ignore the time dependence of the potential  $\phi$ 
in the Einstein equations. 
This approximation holds as long as $\ell/L_H \ll 1$, where $\ell$ is the scale of
 the perturbations we consider. The potential begins to decay when the cosmological constant 
begins to dominate the cosmic expansion. 

At first order, from the Einstein equations we obtain
\begin{eqnarray}
\frac{1}{a^{2}}\phi ,_{ii}&=&\frac{3}{2}\frac{\dot{a}^{2}}{a^{2}}(2\phi + \delta) 
- \frac{\Lambda}{2}\delta,\\
\delta_{ij}v^{j}&=&-2\frac{\dot{a}}{a}(3\dot{a}^{2}
-a^{2}\Lambda)^{-1}\phi _{,i},
\end{eqnarray}
where $\delta = (\rho-\rho_{b})/\rho_{b}$ is the density contrast,  $\rho_{b}$ is the background density, and contraction is taken with  $\delta_{ij}$.

Next, we substitute these solutions into the right-hand sides of Eqs.(16) and (17), 
and retaining terms up to quadratic order in $\phi$,   
the averaged Einstein equations become
\begin{equation}
  \left(\frac{\dot{a}}{a}\right)^2 = \frac{8\pi G}{3}
  \langle{T_{00}}\rangle +
  \frac{1}{a^2} \langle{\phi_{,i}\phi_{,i}}\rangle 
  \frac{\Lambda}{3}\label{eq:t001},
\end{equation}
\begin{equation}
  \frac{\ddot{a}}{a} = -\frac{4\pi G}{3}
  \langle{T_{00} + \rho_b a^2 v^2}\rangle
  - \frac{1}{3a^2}\langle{\phi_{,i}\phi_{,i}}\rangle + \frac{\Lambda}{3},\label{eq:t002}
\end{equation}
where $v^{2}=\delta_{ij}v^{i}v^{j}$. We do not regard these equations as the 
Friedman equations for the averaged FRLW universe, and thus we do not use them 
to interpret the effect of the backreaction on the cosmic expansion. 
This is because these equations reveal that the background density $\rho_b$ is not 
conserved in the comoving volume, and thus $\rho_b$ is not appropriate 
as the mean energy density of the averaged FRLW universe. 

Now we define the conserved mean energy density for the averaged FRLW universe. 
The mean density $\bar{\rho}$ is defined to be the quantity that  satisfies the 
equation
\begin{eqnarray*}
\dot{\bar{\rho}}=-3\frac{\dot{a}}{a}\bar{\rho}.
\end{eqnarray*}
It is straightforward to derive the mean density using the energy conservation law
\begin{equation}
\bar{\rho}\equiv \langle{T_{00}}\rangle 
  +\frac{5}{12\pi Ga^{2}} \langle{\phi_{,i}\phi_{,i}}\rangle 
   +\frac{a\Omega_{\Lambda}}{24\pi G\Omega_{m}} \langle{\phi_{,i}\phi_{,i}}\rangle.
\end{equation}
Here, Eq.(21) has been used to rewrite the term containing $\langle{v^2}\rangle$ in terms of $\langle{\phi_{,i}\phi_{,i}}\rangle$,  
$\Omega_m = \rho_b/\rho_{cr}$, and $\Omega_\Lambda =\Lambda/3 H^2_0$, where  
$H_{0}$ and $\rho_{cr}=H^2_0/8\pi G $ are the Hubble parameter and critical density, 
respectively.

We rewrite Eqs.(\ref{eq:t001}) and (\ref{eq:t002}) in terms of  
the mean density $\bar{\rho}$:
\begin{eqnarray}
\left(\frac{\dot{a}}{a}\right)^{2}
&=&\frac{8\pi G}{3}\bar{\rho}+\frac{\Lambda}{3}-\frac{1}{9}\left(\frac{1}{a^{2}}
+\frac{\Omega_{\Lambda}}{\Omega_{m}}a\right) \langle{\phi_{,i}\phi_{,i}}\rangle, 
\\
\frac{\ddot{a}}{a}&=&-\frac{4\pi G}{3}\bar{\rho} + \frac{\Lambda}{3} 
   -\frac{a\Omega_{\Lambda}}{6\Omega_{m}} \langle{\phi_{,i}\phi_{,i}}\rangle.
\end{eqnarray}
These are the Friedmann equations for the averaged FLRW model of a locally inhomogeneous 
universe. The third terms on the R.H.S. of the above equations 
are interpreted as the non-linear backreaction due to local inhomogeneities.
There are two types of backreaction. One is proportional to $a^{-2}$, and hence 
behaves as a small positive curvature term in the Friedmann equation. 
This is the same as in the Einstein-de Sitter case\cite{KAF}. The other is 
proportional to $a$, and hence behaves as a phantom fluid in this sense. However, this term acts to decelerate the cosmic expansion, as shown in Eq.(26). 
The backreaction induced by the cosmological constant has strange properties. 
If we define an effective energy density $\rho^{(in)}_\Lambda$ and 
effective pressure $P^{(in)}_\Lambda$ of the backreaction, then we find
\begin{eqnarray}
\rho^{(in)}_\Lambda = -\frac{\Omega_{\Lambda}a}{24\pi G\Omega_{m}}\langle{\phi_{,i}\phi_{,i}}\rangle,\\
P^{(in)}_\Lambda =\frac{\Omega_{\Lambda}a}{18\pi G\Omega_{m}}\langle{\phi_{,i}\phi_{,i}}\rangle.
\end{eqnarray}
Thus, the equation of state is $P^{(in)}_\Lambda= -\frac{4}{3} \rho^{(in)}_\Lambda$.  

When the cosmological constant begins to dominate the expansion, the potential begins to 
decay. Thus the backreaction also decays. Note that we have ignored the time 
dependence of the potential when averaging the Einstein equation. 
As discussed above, this is allowed as long as we consider perturbations with 
scales much smaller than the horizon scale.
In fact, we can evaluate the backreaction using the power spectrum $P^\phi(k)$ of 
the gravitational potential as
\begin{equation}
\langle{\phi_{,i}\phi_{,i}}\rangle =\int\frac{d^{3}k}{(2\pi)^3}k^{2}P^{\phi}(k).
\end{equation}
Employing a realistic $\Lambda$CDM model for the power spectrum\cite{PD}, 
we have confirmed that changing the minimum momentum in the integration from $H_0$ to $0.1 H_0$ 
results in a change in the value of $\langle{\phi_{,i}\phi_{,i}}\rangle$  that is less than $0.3\%$.  

We have found that the gravitational interaction between local inhomogeneities and 
the cosmological constant induces a peculiar backreaction which appears in or near the epoch of 
the structure formation only. Its effects on the global expansion and the distance-redshift 
relation are in general very small. 
For example, the backreaction changes the luminosity distance from the present 
to $z=1000$ by less than $0.001\%$ from the standard distance in the  $\Lambda$CDM 
model with $\Omega_m=0.3$ and $\Omega_\Lambda=0.7$. However, it is interesting that the effect of the backreaction changes the observational effect if we properly take 
into account the large scale perturbations which we have ignored in this letter.   

Also, the local inhomogeneities are important 
in cosmological observations. In fact, the scattering of light rays from distant sources 
by local inhomogeneities causes a redshift-dependent dispersion in the distance-redshift 
relation which will be essential to properly interpret cosmological observations 
such as the magnitude-redshift 
relation of Type Ia supernpovae\cite{HF} and the evolution of the quasar luminosity function.

\section*{Acknowledgments}
The authors would like to thank M. Kasai and H. Asada for 
discussions. This work was partly supported by the COE program at Tohoku University


\begin{thebibliography}{99}
  
\bibitem{WMAP}D.~N.~ Spergel et al., \JL{Astrophys.\ J.\ Suppl.,148,2003,175}.
\bibitem{SDSS}M.~Tegmark et al., \AJ{606,2004,702}.
\bibitem{Futamase88}T. Futamase, \PRL{61,1988,2175}.
\bibitem{Futamase89}T. Futamase, \JL{Mon.\ Not.\ R.\ Astron.\ Soc.,237,1989,187}.
\bibitem{Futamase96}T. Futamase, \PRD{53,1996,681}.
\bibitem{Kasai92}M. Kasai, \PRL{69,1992,2330}.
\bibitem{Kasai93}M. Kasai, \PRD{47,1993,3214}.
\bibitem{Kasai95}M. Kasai, \PRD{52,1995,5605}.
\bibitem{HF}T. Hamana and T. Futamase, \JL{Astrophys.\ J.,534,2000,29}.
\bibitem{KAF}M. Kasai, H. Asada and T. Fuatmase, \PTP{115, 2006, 827}. 
\bibitem{Russ96}H. Russ, M. Morita, M. Kasai and G. B\"orner,  \PRD{53,1996,6881}. 
\bibitem{Russ97}H. Russ, M. H. Soffel, M. Kasai and G. B\"orner,  \PRD{56,1997,2044}.
\bibitem{SA}M. Shibata and H. Asada, \PTP{94,1995,11}.
\bibitem{AK}H. Asada and M. Kasai, \PRD{59,1999,0123515}. 
\bibitem{Riess}A. Riess et al., \JL{Astron.\ J.,116,1998,1009}. 
\bibitem{Perlmutter}S. Perlmutter et al., \JL{Nature,391,1998,391}. 
\bibitem{FGC}P. Fosalba, E. Gaztanaga and F. J. Castander,  \AJ{597,2003,L89}.
\bibitem{BC}S. Boughn and R. Crittenden, \JL{Nature,427,2004,45}.
\bibitem{Bahcall}N. Bahcall, J. Ostriker, S. Perlmutter and  P. Steinhardt, \JL{Science,284,1999,1481}. 
\bibitem{Buchert00}T. Buchert, \JL{Gen. Rel. Grav., 32, 2000, 105}.
\bibitem{Buchert06}T. Buchert, \JL{Astron. Astrophs., 454, 2006, 415}.
\bibitem{PR}P. J. E. Peebles and B. Ratra, \JL{Rev.\ Mod.\ Phys.,75,2003,559}.
\bibitem{Rasanen}S. Rasanen, \JL{J. Cosmol. Astropart. Phys.,02,2004,003}. 
\bibitem{Kolb05}E. W. Kolb, S. Matarrese, A. Notari and A. Riotto,  \PRD{71,2005,023524}.
\bibitem{IW}A. Ishibashi and R. M. Wald, \JL{Class. Quant. Grav., 23,2006,235}.
\bibitem{HS}M. Hirata and U. Seljak, \PRD{72,2005, 083501}.
\bibitem{PD}J. A. Peacock and S. J. Dodds, \JL{Mon.\ Not.\ R.\ Astron.\ Soc.,280,1996,L19}.



\end{thebibliography}
\end{document}